\documentclass[preprint,aps]{revtex4}
\usepackage{amssymb,epsf}
\usepackage{epsfig}
\usepackage{epstopdf}
\usepackage{graphicx}
\usepackage{color}

\newcommand{\tcr}{\textcolor{red}}

\begin{document}

\title{Eguchi-Hanson like space-times in $F(R)$ gravity}
\author{S. H. Hendi$^{1,2}$ \footnote{E-mail: hendi2004@gmail.com \& hendi@shirazu.ac.ir}, R. B. Mann$^{3}$
\footnote{E-mail: rbmann@sciborg.uwaterloo.ca}, N. Riazi$^{1}$
\footnote{E-mail: riazi@physics.susc.ac.ir} and B. Eslam
Panah$^{4}$} \affiliation{$^1$ Physics Department and Biruni
Observatory, College of Sciences, Shiraz
University, Shiraz 71454, Iran\\
$^2$ Research Institute for Astrophysics and Astronomy of Maragha (RIAAM),
P.O. Box 55134-441, Maragha, Iran\\
$^3$ Department of Physics, University of Waterloo, 200 University Avenue
West, Waterloo, Ontario, Canada N2L 3G1\\
$^4$ Department of Physics, University of Tabriz, Tabriz, Iran}

\begin{abstract}
We consider a model of $F(R)$ gravity in which exponential and
power corrections to Einstein-$\Lambda$ gravity are included. We
show that this model has 4-dimensional Eguchi-Hanson type
instanton solutions in Euclidean space. We then seek solutions to
the five dimensional equations for which space-time contains a
hypersurface corresponding to the Eguchi-Hanson space. We obtain
analytic solutions of the $F(R)$ gravitational field equations,
and by assuming certain relationships between the model parameters
and integration constants, find several classes of exact
solutions. Finally, we investigate the asymptotic behavior of the
solutions and compute the second derivative of the $F(R)$ function
with respect to the Ricci scalar to confirm Dolgov-Kawasaki
stability.
\end{abstract}

\maketitle

\section{Introduction}

In recent years, much work has been done in order to bring us more
information on extra dimensions. In this domain, investigation of black
objects, soliton solutions and instantons are of particular importance, and could
play a key role in opening a window to extra dimensions.

In four dimensions, it is well known that some of the static black
hole solutions described by the Majumdar-Papapetrou solution
\cite{Majumdar} can be extended to multi-black objects in higher
dimensions \cite{Myers}. The four dimensional Riemannian manifolds
for gravitational instantons can be asymptotically flat,
asymptotically locally Euclidean, asymptotically locally flat or
compact without boundary. For example, Hawking's interpretation of
the Taub-NUT solution \cite{HawPLA} is an example of
asymptotically locally flat space. The simplest nontrivial example
of asymptotically locally Euclidean spaces is the metric of
Eguchi-Hanson \cite{EHPLB}. Asymptotically locally Euclidean
instantons have been found explicitly by Gibbons and Hawking
\cite{GibHaw} and they are known implicitly through the work of
Hitchin \cite{Hitchin}. The complex projective space $\mathbb{CP}^{2}$ is
an example of compact, anti self-dual instanton solving Einstein's
equation with a cosmological constant term \cite{EgFrPRL} .

Some what more recently, new soliton solutions with interesting
properties were discovered a few years ago in 5-dimensional
Einstein gravity \cite{ClarkMann, Copsey}. These solutions
resemble the Eguchi-Hanson metrics \cite{EHPLB} with $AdS/Z_{p}$
asymptotics. Eguchi-Hanson metrics were originally obtained
as solutions to the 4-dimensional Euclidean Einstein equations
\cite{EH} and have the form
\begin{equation}
ds^{2}=\frac{dr^{2}}{1-\frac{\alpha
^{4}}{r^{4}}}+\frac{r^{2}}{4}\left( \left( d\psi+\cos \theta
d\varphi \right)^2  + d\theta^2 + \sin^2\theta d\varphi^2 \right)
\label{EHoriginal}
\end{equation}
where $\alpha$ is a constant of integration. The Riemann curvature
tensor of these solutions is self-dual, obeying the relation
$\epsilon_{\alpha\beta}^{\phantom{\alpha\beta}\mu\nu}R_{\mu\nu\rho\tau}
=R_{\alpha\beta\rho\tau}$.

Once we entertain the notion of extra dimensions, we need not
impose the asymptotic flatness condition, spherical topology of
the horizon and other four dimensional restrictions. In fact,
higher dimensional solutions admit a variety of asymptotic
structures and horizon topologies that have more interesting
properties and richer structure than four-dimensional ones. For
example, there have been many interesting results in Kaluza-Klein
theory \cite{Kaluza} with compact extra dimensions as well as in
solutions making use of Eguchi-Hanson and Taub-NUT spaces
\cite{EH,NUT} with nontrivial topology and structure.

From a geometrical point of view, one of the simplest
modifications of the gravitational interaction to higher order is
$F(R)$ gravity, whose action is an arbitrary function of the
curvature scalar $R$ \cite{FR,Cognola2008,Sotiriou,HendiPLB}. A
key motivation for considering this class of theories has to do
with addressing a number of cosmological problems. These include
the power law early-time inflation
\cite{Starobinsky1980,Bamba2008}, late-time cosmic accelerated
expansion \cite{Bamba2008,Carroll2004,Dombriz2006,Fay2007} and the
singularity problem arising in the strong gravity regime
\cite{Abdalla,Briscese,Noj046006,Bamba,Kobayashi2008}. Furthermore
one can find an explanation of the hierarchy problem
\cite{Cognola2006}, the four cosmological phases \cite{Nojiri2006}
and the rotation curves of spiral galaxies \cite{Capozziello2006}
within $F(R)$ gravity.

Following the method of \cite{Hendi} (which is concerned with a
known class of the $F(R)$ theory instead of Einstein gravity) the
main scope of this work is to present some results on
Eguchi-Hanson like metrics and investigate their interesting
properties. Although many of the solutions we obtain have
unphysical properties, we find generalizations of the
Eguchi-Hanson instanton in 4 dimensions, and a new soliton
solution in $5$ dimensions.

In contrast with general relativity,  $F(R)$ gravity may
be considered intrinsically unstable. This difference is due to
the fact that the field equations in general relativity are of
second order and therefore their trace gives an algebraic equation
for Ricci scalar, but the field equations of $F(R)$ gravity are of
fourth order and so their trace gives a dynamical equation for
Ricci scalar. \tcr{It has been noted that if $F(R ) \neq 0$ and  $dF/dR=0$ then no stable 
ground state existed \cite{JSchmidt}.  For the solutions we obtain we find that   $F(R ) = dF/dR=0$.
Furthermore,}
it has been shown that the effective mass of the
dynamical field of Ricci scalar is related to the second
derivative of the $F(R)$ with respect to $R$ \cite{Dolgov1}.
Therefore, in order for the dynamical field to be stable its
effective mass must be positive, a requirement usually referred to
as Dolgov-Kawasaki stability criterion. The Dolgov-Kawasaki
instability criterion, which has found in the metric version of
$F(R)$ theories, is sufficiently strong to veto some models
\cite{Dolgov2}.

The outline of our paper is as follows. In section \ref{FieldEq}
we present a short review of field equation of $d$-dimensional
$F(R)$ gravity. This field equation is then solved for
4-dimensional Eguchi-Hanson space in Sec. \ref{E-H space}. We then
generalize it to 5-dimensional Eguchi-Hanson like space-time and
obtain a new soliton solution. Conclusions are drawn in the last
section, and we present more solutions using the same approach in
the appendix.

\section{ Basic Field Equations \label{FieldEq}}

We start from the following action of pure general $F(R)$ gravity
\begin{equation}
\mathcal{I}_{G}=-\frac{1}{16\pi }\int d^{d}x\sqrt{-g}F(R),  \label{Action}
\end{equation}
Variation of Eq. (\ref{Action}) with respect to metric $g_{\mu \nu }$, leads
to the field equation of $F(R)$ gravity
\begin{equation}
R_{\mu \nu }F_{R}-\nabla _{\mu }\nabla _{\nu }F_{R}+\left( \Box
F_{R}-\frac{1}{2}F(R)\right) g_{\mu \nu }=0,  \label{FE}
\end{equation}
where ${R}_{\mu \nu }$ is the Ricci tensor and $F_{R}\equiv dF(R)/dR$. We
consider a recently proposed \cite{Hendi} $F(R)$ model
\begin{equation}
F(R)=R-2\Lambda -\lambda \exp (-\xi R)+\kappa R^{n}  \label{FR}
\end{equation}
in which exponential and power corrections to Einstein-$\Lambda$
gravity are included. This choice of $F(R)$ gravity has some
interesting properties such as providing charged solutions from
pure $F(R)$ gravity \cite{Hendi}.

Viable modifications to Einstein gravity must pass all sorts of
empirical tests, from the large scale structure of the universe to
galaxy and cluster dynamics to solar system tests. One of the
outstanding questions in $F(R)$ gravity is whether it is
consistent with solar system tests or not. When the correction
term to Einstein gravity is of the exponential form
\cite{Cognola2008,EXPform}, it can be shown that there is no
conflict with solar system tests and that the high curvature
condition is satisfied \cite{Zhang}. Furthermore, the solutions of
this model are virtually indistinguishable from those in general
relativity, up to a change in Newton's constant \cite{Zhang}.

In addition, one can choose $R^{2}$ with exponential corrections to Einstein
gravity \cite{EXPform,Starobinsky} to explain inflation. By adjusting the
free parameters, this model can satisfy the high curvature condition, early
universe inflation, stability and the late time acceleration \cite{Zhang}.
Also, in order to resolve the singularity problem arising in the strong
gravity regime, Kobayashi and Maeda \cite{Kobayashi2008} have considered a
higher curvature correction proportional to $R^{n}$ where $n>1$.

Motivated by the above, we seek instanton solutions to $F(R)$
gravity using (\ref{FR}) for 4-dimensional Eguchi-Hanson space and
soliton solutions for Eguchi-Hanson like $(4+1)$ -dimensional
space-time.

\section{Generalized Eguchi-Hanson Instantons \label{E-H space}}

We consider first the $4$-dimensional Eguchi-Hanson type metric
ansatz
\begin{equation}
ds^{2}=\frac{dr^{2}}{g(r)}+r^{2}\left( \sigma _{x}^{2}+\sigma
_{y}^{2}\right) +r^{2}g(r)\sigma _{z}^{2},  \label{EHSMet}
\end{equation}
where the differential one forms ${\sigma }_{i}$ can expressed in terms of
Euler angles ${\theta }$, ${\phi }$ and ${\psi }$
\begin{eqnarray}
\sigma _{x} &=&\frac{1}{2}\left( \sin \psi d\theta -\sin \theta \cos \psi
d\varphi \right) ,  \nonumber \\
\sigma _{y} &=&\frac{1}{2}\left( -\cos \psi d\theta -\sin \theta \sin \psi
d\varphi \right) ,  \nonumber \\
\sigma _{z} &=&\frac{1}{2}\left( d\psi +\cos \theta d\varphi \right) .
\end{eqnarray}
For $F(R)=R$, it is well known that the solution to the Euclidean Einstein
equations is \cite{EHPLB}
\begin{equation}
g(r)=1-\frac{\alpha ^{4}}{r^{4}}  \label{EHmet}
\end{equation}
where $\alpha $ is constant. For $\alpha=0$ the constant-$r$
surfaces of the metric (\ref{EHSMet}) have the geometry $S^3$,
whereas for $\alpha\neq0$ the geometry must $S^3/\mathbb{Z}_2$ to
avoid singularities.  Note that the metric $\sigma _{x}^{2}+\sigma
_{y}^{2}$ yields the geometry of a 2-sphere of radius $1/2$.

We wish to obtain general solutions using the
ansatz (\ref{EHSMet}) for the mentioned model of $F(R)$ theory, and
investigate their geometrical properties.

Since the metric is purely Euclidean with no time-like direction,
any solutions we obtain cannot be interpreted as black holes. If
the function $g(r)$ vanishes at $r=r_0$ then the metric will have
a singularity unless $\psi$ is appropriately periodically
identified with period
\begin{equation}
\Delta \psi = \frac{16\pi}{r_0 g^\prime(r_0)}  \label{psiper}
\end{equation}
where $g^\prime \equiv dg/dr$.

Using Eqs. (\ref{EHSMet}) and (\ref{FE}), we find the solution
\begin{equation}
g(r)=1-\frac{\chi r^{2}}{24}+\frac{q^{2}}{r^{2}}+\frac{a}{r^{4}},
\label{faI}
\end{equation}
\begin{eqnarray}
\xi &=&\frac{\Pi }{\Omega },  \nonumber \\
\lambda &=&-\Omega e^{\frac{\chi \Pi }{\Omega }},  \label{faIparameters}
\end{eqnarray}
for the metric function, where $\Pi =n\kappa \chi ^{n-1}+1$, $\Omega
=2\Lambda -\kappa \chi ^{n}-\chi $, and $\chi $, $q$ and $a$ are integration
constants and $\xi $, $\beta $, $\kappa $ and $n$ are free parameters. In
order to study the geometrical structure of this solution, we first look for
the essential singularity(ies). The Ricci scalar and the Kretschmann scalar
are
\begin{equation}
R=\chi,
\end{equation}
and
\begin{equation}
R_{\mu \nu \rho \sigma }R^{\mu \nu \rho \sigma
}=\frac{384a^{2}}{r^{12}}+\frac{384aq^{2}}{r^{10}}+\frac{128q^{4}}{r^{8}}+\frac{\chi
^{2}}{3}, \label{KreI}
\end{equation}
which means that the Kretschmann scalar (\ref{KreI}) diverges at
$r=0$, is finite for $r\neq 0$ and approaches $\frac{\chi
^{2}}{3}$ as $r\longrightarrow \infty $.

\begin{figure}[tbp]
\epsfxsize=10cm \centerline{\epsffile{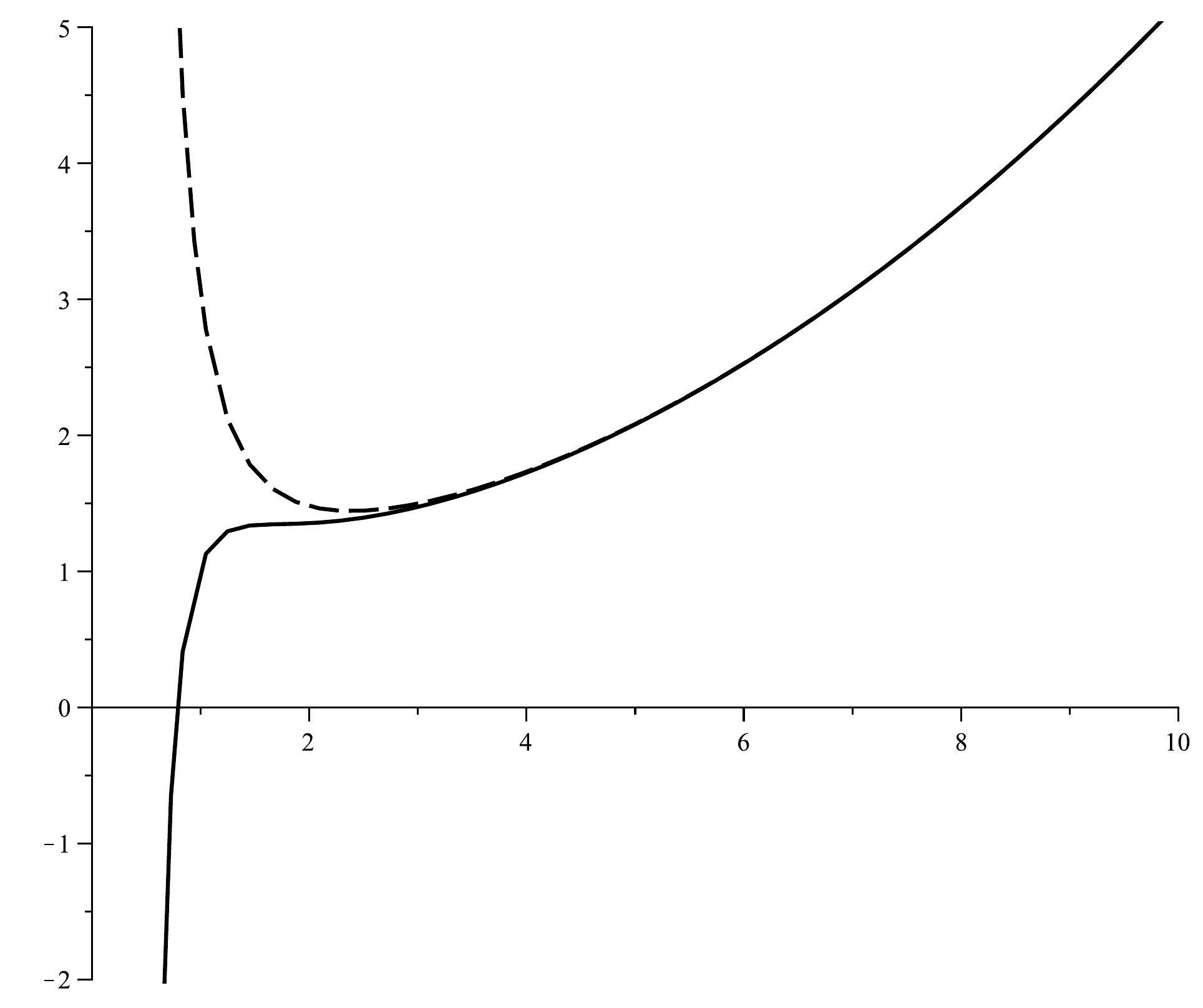}}
\caption{ Eq. (\protect\ref{faI}): $g(r)$ versus $r$ for $\protect\chi =-1$,
$q=1$, and $a=-1$ (solid line) and $a=1$ (dashed line).}
\label{Fig1}
\end{figure}
\begin{figure}[tbp]
\epsfxsize=10cm \centerline{\epsffile{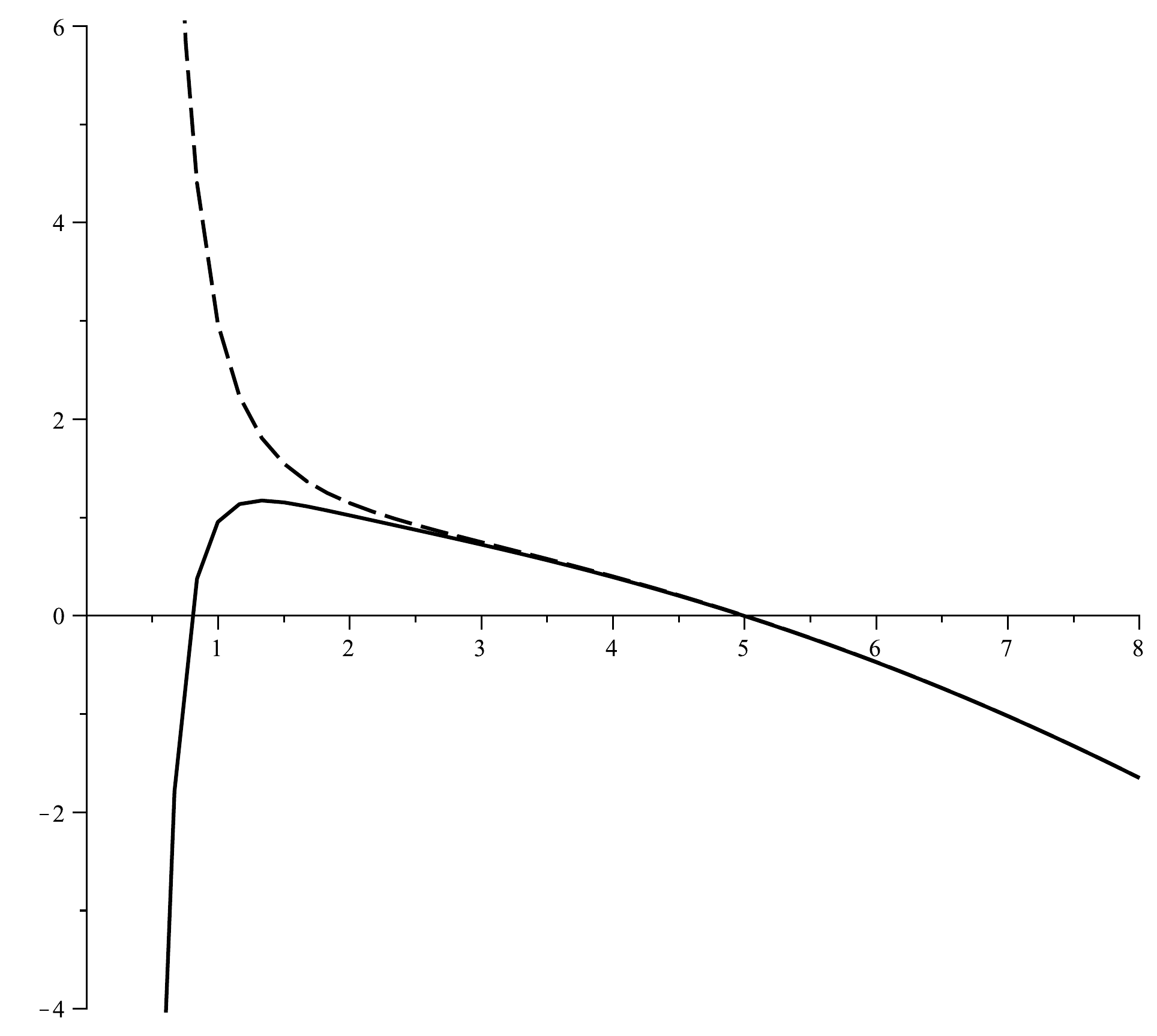}} \caption{ Eq.
(\protect\ref{faI}): $g(r)$ versus $r$ for $\protect\chi =1$,
$q=1$, and $a=-1$ (solid line) and $a=1$ (dashed line).}
\label{Fig2}
\end{figure}

For $\chi <0$ and positive $a$ this solution has a singularity at $r=0$.
However for $\chi <0$ and negative $a$ the function $g(r)$ vanishes at a
finite value of $r$ we denote by $r_{0}$. The space is geodesically complete
provided the coordinate $\psi $ has period
\[
\frac{64\pi r_{0}^{2}}{3|\chi |r_{0}^{4}+8q^{2}+16r_{0}^{2}}
\]
where
\[
\frac{3|\chi |r_{0}^{6}}{24}+r_{0}^{4}+q^{2}r_{0}^{2}-|a|=0
\]
Figure \ref{Fig1} illustrates that for negative $a$, the function $g(r)$
vanishes at a certain value of $r$; this solution is like a ``bubble" in
Euclidean space or a kind of soliton.

For $\chi >0$ and positive $a$ {the function $g(r)$ }has a real
root at $r_{0}$, becoming negative for larger values of $r$. This
space has a naked singularity at $r=0$. However if $a$ is also
negative then the metric function $g(r)$ has two roots, $r_{<}$
and $r_{>}$, which delineate the range of $r$ (see Fig.
\ref{Fig2}). The solution will be free of singularities provided
the period of $\psi $ at $r_{<}$ is the same as that at $r_{>}$.
We find that there are no real values of the parameters for which
this situation holds, and so conclude that all $\chi >0$ solutions
are singular.

This solution is problematic insofar as the circle described by
$\psi $ is growing exponentially relative to the 2-sphere as
$r\rightarrow \infty $. The reason is that $g(r)$ is growing like
$r^{2}$ for large $r$, so the $\psi $-circle is growing much
faster than the $2$-sphere for large $r$. This will not happen if
$\chi =0$, which implies $\xi =\frac{1}{2\Lambda }$ and $\lambda
=-2\Lambda $.

We therefore conclude that $\chi =0$ and $a<0$ yields the only physically
reasonable class of instanton solutions, whose metric is
\begin{equation}
ds^{2}=\frac{dr^{2}}{1+\frac{q^{2}}{r^{2}}-\frac{\alpha^4}{r^{4}}}
+r^{2}\left( \sigma _{x}^{2}+\sigma _{y}^{2}\right)
+r^{2}\left(1+\frac{q^{2}
}{r^{2}}-\frac{\alpha^4}{r^{4}}\right)\sigma _{z}^{2},
\label{EHSMet2}
\end{equation}
where we have set $a=-\alpha ^{4}$. We find that the metric
function vanishes at $r_{0}=\frac{\sqrt{2}}{2}(\sqrt{q^{4}+4\alpha
^{4}}-q^{2})$ and that
\begin{equation}
\Delta \psi ={4\pi }\left( 1-\frac{q^{2}}{\sqrt{q^{4}+4\alpha ^{4}}}\right)
\label{psiper2}
\end{equation}
generalizing the Eguchi-Hanson metric (\ref{EHmet}) to $F(R)$
gravity. Note that the class of metrics (\ref{EHSMet2})
cannot be obtained by Wick rotation of some coordinate in a
corresponding Lorentzian-signature space-time.  

\begin{figure}[tbp]
\epsfxsize=10cm \centerline{\epsffile{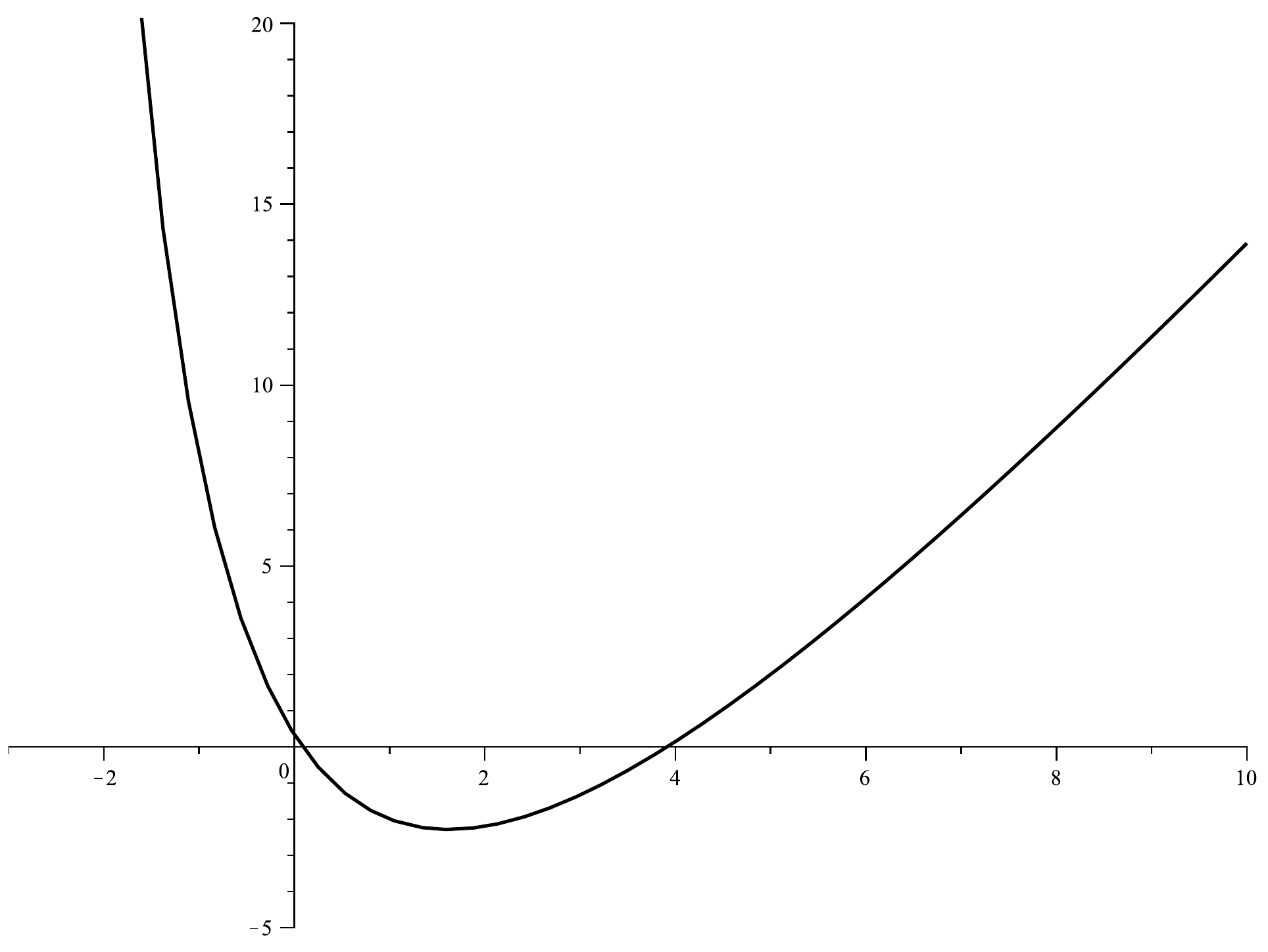}} \caption{ Eq.
(\protect\ref{EHFRR}): $F_{RR}$ versus $\protect\kappa$ for
$\protect\chi =-1$, $\Lambda =1$ and $n=3$.} \label{FRREHspace}
\end{figure}

Although we have not included a time-like coordinate yet, the
$F(R)$ model can be checked for stability. In order to check the
stability condition (Dolgov-Kawasaki stability), we obtain the
second derivative of the $F(R)$ function with respect to Ricci
scalar
\begin{equation}
F_{RR}=\frac{\Pi ^{2}}{\Omega }+n(n-1)\kappa \chi ^{n-2},  \label{EHFRR}
\end{equation}
which shows that this model is stable provided the free parameters of the
model are chosen suitably (see Fig. \ref{FRREHspace} for more details). We
should note that for $\chi =0$, $F_{RR}$ reduces to
\begin{equation}
F_{RR}=\frac{1}{2\Lambda }.
\end{equation}
which confirms that this solution is stable for $\Lambda >0$.

\section{5-dimensional Soliton Solutions \label{E-H like space-time}}

Here, we include the time coordinate and begin with the following
$5$-dimensional ansatz for the metric
\begin{equation}
ds^{2}=-h(r)dt^{2}+\frac{dr^{2}}{h(r)g(r)}+r^{2}\left( \sigma
_{x}^{2}+\sigma _{y}^{2}\right) +r^{2}g(r)\sigma _{z}^{2}.  \label{E-Hlike}
\end{equation}
In this metric, the hypersurface $t=$constant corresponds to Eguchi-Hanson
space when we fix $h(r)=1$ and set $g(r)$ as in (\ref{EHmet}). For $g(r)=1$,
Eq. (\ref{E-Hlike}) reduces to
\begin{equation}
ds^{2}=-h(r)dt^{2}+\frac{dr^{2}}{h(r)}+\frac{r^{2}}{4}\left( d\theta
^{2}+d\phi ^{2}+d\psi ^{2}\right) +\frac{r^{2}}{2}\cos (\theta )d\phi d\psi ,
\end{equation}
which is Schwarzschild-like space-time. When we choose $h(r)=1$,
one can find that $g(r)$ is the same as Eq. (\ref{faI}) with the
same analysis. In this section, we are looking for exact solutions
of the field equations (\ref{FE}) for the Eguchi-Hanson like
metric ansatz (\ref{E-Hlike}) with $F(R)$ given by Eq. (\ref{FR}).

We should note that for any possible solution for the metric
ansatz (\ref{E-Hlike}), one may encounter several cases of
asymptotic structures and singularities of the metric:

\begin{itemize}
\item $h(r)$ flips sign at one or more values of $r$ and $g(r)$ does not
change sign for sufficiently large $r$. These correspond to the usual kinds
of horizons: black hole and cosmological. If $h(r)\longrightarrow 1$ at
large $r $ then there will be no cosmological horizon. There will be a black
hole if $h(r)$ flips sign at small $r$ before $g(r)$ does.

\item $g(r)$ flips sign at one or more values of $r$ and $h(r)$ does not
change sign. At small finite $r>0$ this will be the edge of the soliton (or
bubble), and if $g(r)\longrightarrow 1$ at large $r$ then the space-time
will be a soliton. If $g(r)$ flips sign again for large $r$ then the space
will be compact.

\item $g(r)$ flips sign at one or more values of $r$ and $h(r)$ changes sign
at other values of $r$. If $g(r_{+})=0$ and $h(r_{c})=0$ then this is a
soliton surrounded by a cosmological horizon. If $h(r)$ has multiple roots,
all larger than $r_{+}$, then the soliton is in some kind of black hole that
has multiple horizons; any roots in the range $r<r_{+}$ don't matter because
the soliton covers them up.

\item $g(r)$ and $h(r)$ both flip sign at the same value of $r$. In this
case the coordinate $\psi $ becomes time-like and the space-time
has closed time-like curves (CTCs) beyond this value of $r$
(either larger or smaller, depending on where $g(r)$ and $h(r)$
are both positive).
\end{itemize}

The only way to get a sensible AdS, or dS, flat structure
asymptotically is if $g(r)\longrightarrow 1$ and
$h(r)\longrightarrow \pm r^{2}$ or $h(r)\longrightarrow 1$ for
large $r$ -- anything else gives some other geometry. For example
if $h(r)\longrightarrow \pm r^{p}$ (and $g(r)\longrightarrow 1$),
then the geometry will be asymptotic to a Lifshitz geometry.

\subsubsection{First solution set:}

We find a more interesting class of space-times if we set $\lambda
=\frac{ e^{\xi \chi }\left[ 3\chi -10\Lambda -\kappa \left(
2n-5\right) \chi ^{n} \right] }{2\xi \chi +5}$. Solving the field
equations (\ref{FE}) and (\ref{E-Hlike}) we find
\begin{eqnarray}
g(r) &=&1-\frac{a^4}{r^{4}},  \label{fgcI} \\
h(r) &=&1-\frac{\chi }{20}r^{2}.  \nonumber
\end{eqnarray}

After some algebraic manipulation, we find that the Kretschmann and the
Ricci scalars are
\begin{eqnarray}
R_{\mu \nu \rho \sigma }R^{\mu \nu \rho \sigma } &=&\frac{384a^{8}}{r^{12}}-
\frac{96\chi a^{8}}{5r^{10}}+\frac{9a^{8}\chi ^{2}}{50r^{8}}+\frac{\chi ^{2}
}{10},  \label{KrecI} \\
R &=&\chi ,  \label{RiccI}
\end{eqnarray}
indicating a curvature singularity at $r=0$, and a space-time of constant
curvature as $r\longrightarrow \infty $. The singularity in (\ref{KrecI}) is
not accessible from the space-time, which is free of singularities provided
the period of $\psi$ is appropriately chosen \cite{ClarkMann}. This solution
is the same as the Eguchi-Hanson soliton \cite{ClarkMann,Copsey}, and will
therefore share all of its properties.

\begin{figure}[tbp]
\epsfxsize=10cm \centerline{\epsffile{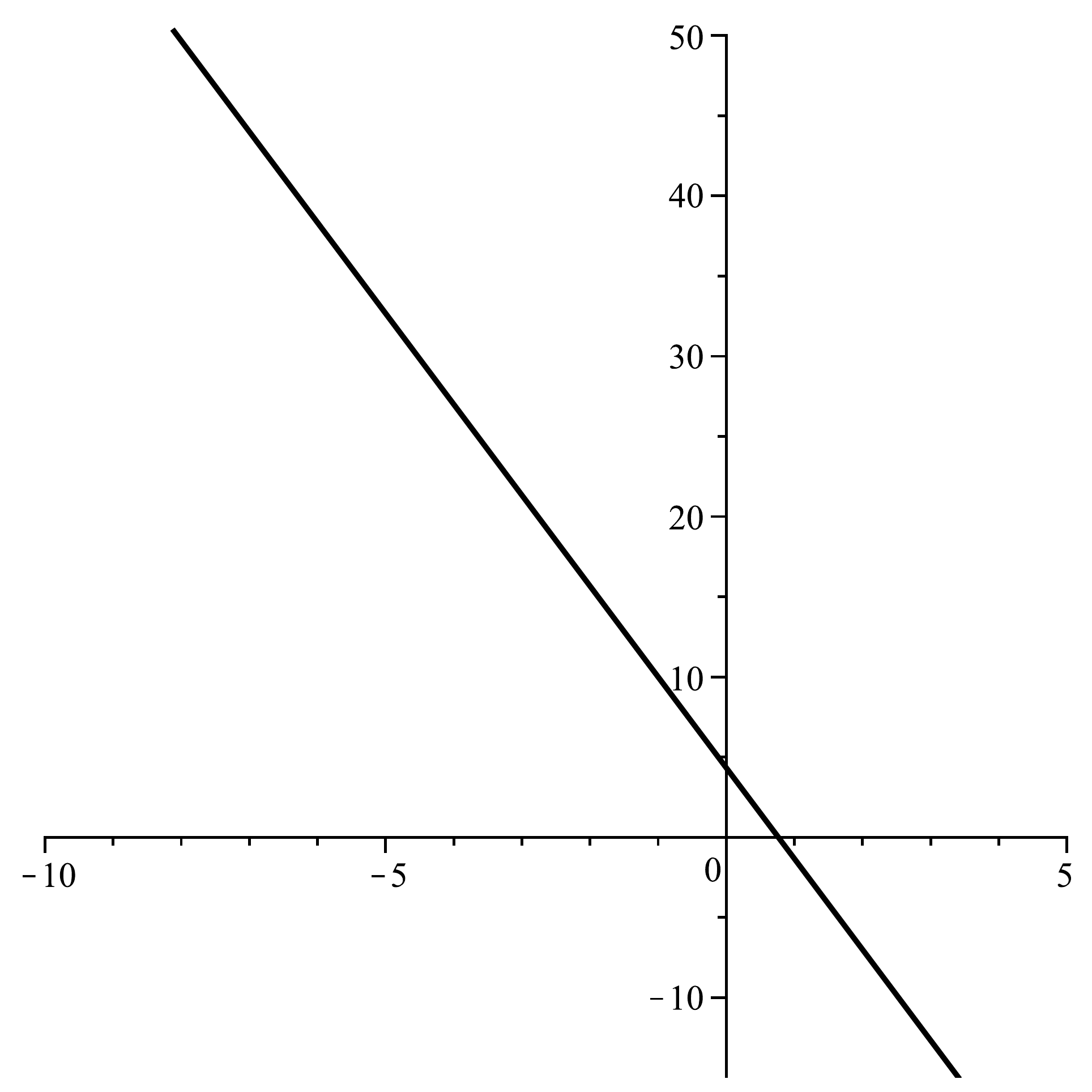}} \caption{ Eq.
(\protect\ref{FRRcI}): $F_{RR}$ versus $\protect\kappa $ for
$\protect\chi =-1$, $\Lambda =1$, $n=3$ and $\protect\xi =1$.}
\label{EHLsol1}
\end{figure}

To investigate Dolgov-Kawasaki stability, we compute derivative of the $F(R)$
function
\begin{equation}
F_{RR}=\frac{\left[ \kappa (2n-5)\chi ^{n}-3\chi +10\Lambda
\right] \xi ^{2}}{2\xi \chi +5}+n(n-1)\kappa \chi ^{n-2},
\label{FRRcI}
\end{equation}
and see that for suitable values of free parameters this model is stable
(see Fig. \ref{EHLsol1}).

\subsubsection{Second solution set:}

Here, we produce another set of solutions with some limitation on the free
parameters of the model. We will see that for a certain choice of parameters
we obtain a new soliton solution that generalizes the Eguchi-Hanson soliton.

Choosing the following parameters
\begin{eqnarray}
n &=&0,  \nonumber \\
\lambda  &=&e^{\frac{-\chi }{\kappa +\chi -2\Lambda }}\left( \kappa +\chi
-2\Lambda \right) ,  \nonumber \\
\xi  &=&\frac{-1}{\left( \kappa +\chi -2\Lambda \right) },
\label{thirdfactor}
\end{eqnarray}
we find a new solution set of the field equation (\ref{FE}) with metric
function (\ref{E-Hlike})
\begin{equation}
\begin{array}{l}
g(r)=1-\frac{5M\chi }{96}-\frac{\chi
r^{2}}{24}-\frac{b}{r^{3}\sqrt{r^{2}-M}}, \\
h(r)=1-\frac{M}{r^{2}},
\end{array}
\label{ThirdSol}
\end{equation}
where $M$, $\chi $ and $b$ are integration constants, and we have
redefined $\kappa -2\Lambda $ as an effective cosmological
constant $\Lambda _{eff}$.

\begin{figure}[tbp]
\epsfxsize=10cm \centerline{\epsffile{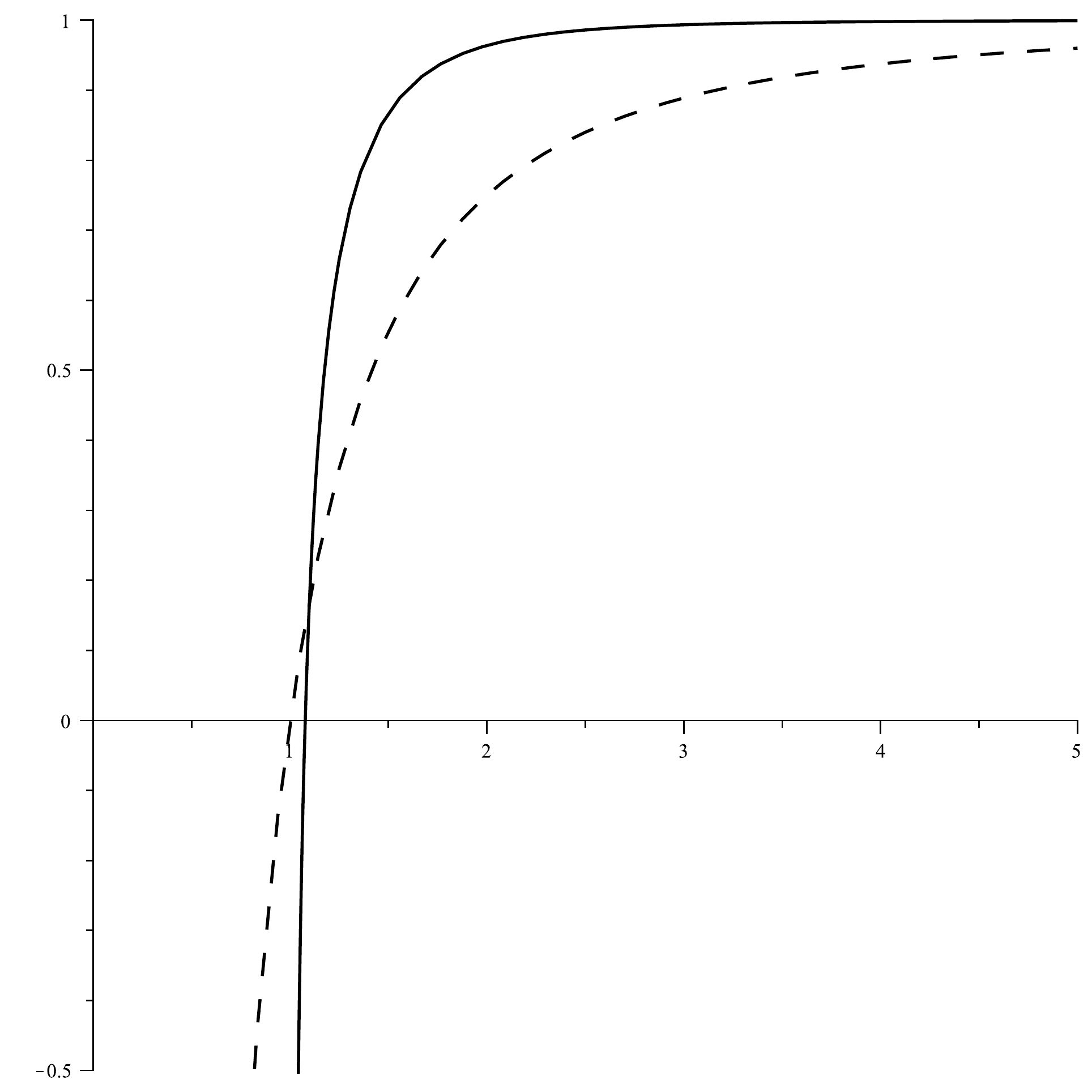}}
\caption{ Eq. (\protect\ref{ThirdSol}): $g(r)$ (solid line) and $h(r)$
(dashed line) versus $r$ for $\protect\chi =0$, $M=1$ and $b=0.5$.}
\label{gh3}
\end{figure}

Unless $\chi =0$ the metric will not have the desired asymptotic
behavior, and so we make this choice, yielding an asymptotically
(locally) flat metric. Note that for positive $M$ this solution is
real for all $r>M$; at $r=M$ it is singular. However this
singularity is excised from the space-time for all $b>0$, since
the function $g(r)$ has a single root at $r>M$, and this root is
larger than the positive root of $h(r)$, as illustrated in figure
\ref{gh3}.

\begin{figure}[tbp]
\epsfxsize=10cm \centerline{\epsffile{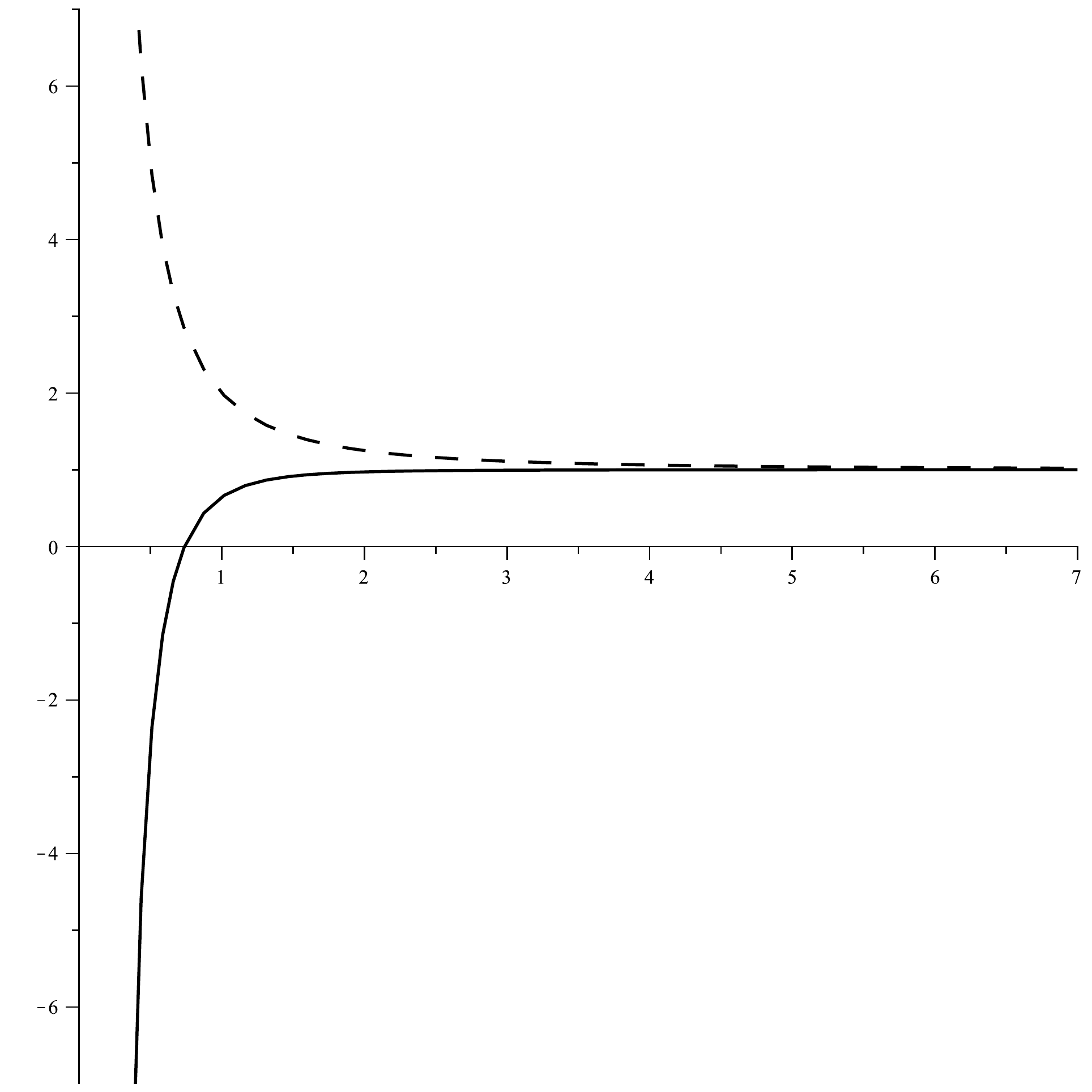}}
\caption{ Eq. (\protect\ref{ThirdSol}): $g(r)$ (solid line) and $h(r)$
(dashed line) versus $r$ for $\protect\chi =0$, $M=-1$ and $b=-0.5$.}
\label{gh33}
\end{figure}

The metric is
\begin{equation}
ds^{2}=-\left( 1-\frac{M}{r^{2}}\right)
dt^{2}+\frac{dr^{2}}{\left( 1-\frac{M }{r^{2}}\right) \left(
1-\frac{b}{r^{3}\sqrt{r^{2}-M}}\right) } +r^{2}\left( \sigma
_{x}^{2}+\sigma _{y}^{2}\right) +r^{2}\left( 1-\frac{b}{
r^{3}\sqrt{r^{2}-M}}\right) \sigma _{z}^{2}  \label{EHMet3}
\end{equation}
and is an interesting new generalization of the EH soliton. The
solution is nonsingular and the quantity $M$ can be of either sign
(see figure \ref{gh33} for the behavior of the metric functions
for $M<0$), with the size of the soliton an increasing function of
$M$.


The soliton radius $r_{s}$ is given by solving a quartic equation, which
yields
\begin{eqnarray}
r_{s\pm } &=&\sqrt{|M|}\left[ \text{sgn}(M)\left(
\frac{1}{4}+\frac{1}{12}
\sqrt{9+6Y^{1/3}-288\sigma ^{8}Y^{-1/3}}\right) \right.   \nonumber \\
&&\quad \left. +\frac{\sqrt{6}}{12}\left( \sqrt{3-Y^{1/3}+48\sigma
^{8}Y^{-1/3}+\frac{9}{\sqrt{9+6Y^{1/3}-288\sigma
^{8}Y^{-1/3}}}}\right) \right] ^{1/2}
\end{eqnarray}
where $\sigma =b^{4}/M^{2}$ and $Y=12\sigma ^{8}\sqrt{768\sigma ^{8}+81}$.
Each of $r_{s\pm }$ are increasing functions of $\sigma $, growing linearly
with $\sigma $ for $\sigma $ sufficiently large.

To ensure regularity of the soliton, the periodicity of $\psi $ must be
chosen so that
\begin{equation}
\Delta \psi =\frac{8\pi r_s\sqrt{r_{s}^{2}-M}}{\left( 4r_{s}^{2}-3M\right)
}=\frac{4\pi }{p}  \label{Period}
\end{equation}
where $p$ is an integer, the latter equality following from the
elimination of string singularities at the poles. Note that if
$M=0$ then $p=2$, consistent with the Eguchi-Hanson instanton.

The above relation constrains the value of $\sigma $ for any
integer $p$. It is straightforward to show that solutions exist
for all integer values of $p>1$. If $M>0$ then the solution
$\sigma =0.720202$ for $p=2$, decreasing as $p$ increases, whereas
if $M<0$ then the solution $\sigma =1.090507733$ for $p=2$,
decreasing toward unity as $p$ increases.

The ADM mass of this solution can be calculated using, for
example, the background subtraction method, and is easily shown to
be ${\cal M} =  3\pi M/8 p$.

\subsubsection{Third solution set:}

Here we present second solution of Eqs. (\ref{FE}) and
(\ref{E-Hlike}) with $ R=\chi $ as
\begin{eqnarray}
g(r) &=&\frac{-3\chi
}{10C}-\frac{a^{4}}{r^{4}}+\frac{\tilde b}{r^{4}\sqrt{Cr^{2}+6}
}+\frac{4\left( 10C+3\chi \right) }{5C^{2}r^{2}},  \nonumber \\
h(r) &=&1+\frac{C}{6}r^{2},  \label{SecondSol}
\end{eqnarray}
where we should adjust
\begin{eqnarray}
\kappa &=&0,  \nonumber \\
\lambda &=&e^{\frac{-\chi }{\chi -2\Lambda }}\left( \chi -2\Lambda \right) ,
\nonumber \\
\xi &=&\frac{-1}{\left( \chi -2\Lambda \right) }  \label{secondfactor}
\end{eqnarray}
which confirm that considering Eq. (\ref{SecondSol}), leads to
vanishing the $R^{n}$ correction. For this solution to have
physically reasonable asymptotic properties, we must set $\chi
=-10C/3$, yielding
\begin{eqnarray}
g(r) &=&1-\frac{a^{4}}{r^{4}}+\frac{b}{r^{4}\sqrt{r^{2}/\ell^2+1}},  \nonumber \\
h(r) &=&1+\frac{r^{2}}{\ell^2},
\end{eqnarray}
where to have real solutions, we have set $C=6/\ell^2>0$, and we
have rescaled $\tilde b \to b$ for convenience. If $b>0$ then the
metric function $g(r)$ has one  real root at $r=r_s>0$, given by
the square root of the real solution to the equation
$$
(r^4_s+b)^2 = a^8(r_s^2/\ell^2 +1)
$$
which is a quartic equation in $\sqrt{r_s}$ . For large $r$ this
space-time is asymptotic to the Eguchi-Hanson soliton
\cite{ClarkMann}. The curvature diverges at $r=0$ (and hence not
at a point located within the space-time).  For large values of
$r$, one obtains
\begin{equation}
\lim_{r\longrightarrow \infty }R_{\alpha \beta \mu \nu }R^{\alpha
\beta \mu \nu }=\frac{40}{l^4}.  \label{RRsecond}
\end{equation}

Regularity in the ($r$, $\psi$) section and also elimination of
string singularities at the north and south poles implies that
\begin{equation}
\Delta \psi =\frac{8\pi r_{s}^{2} \ell\sqrt{r_{s}^{2}+l^2}}{
5r_{s}^{4}+4r_{s}^{2} l^2-a^4}=\frac{4\pi }{p} \label{Period3}
\end{equation}

Here, we analyze the Dolgov-Kawasaki stability. It is easy to show that
\begin{equation}
F_{RR}=\frac{l^2}{20 +2\Lambda l^2 }, \label{FRRsecond}
\end{equation}
and therefore we conclude that this solution is stable for
$\Lambda>-\frac{10}{l^2}$.

It is straightforward to show using the methods of Ref.
\cite{Das:2000cu} that the mass of this solution is ${\cal M} =
-\pi a^4/8\ell^2 p$, the same value as for the Eguchi-Hanson
soliton \cite{ClarkMann}, apart from a constant factor due to the
Casimir energy.

\section{Conclusions}

In this work, we have considered a kind of well-known $F(R)$
gravity in Eguchi-Hanson space and Eguchi-Hanson like space-time.
We have shown that, in $4$-dimensional Eguchi-Hanson type
Euclidean metric, the solutions can be interpreted as stable
instantons.

Upon including the time coordinate we found two distinct
generalized $5$-dimensional Eguchi-Hanson space-times in $F(R)$
gravity, one that is asymptotically flat and another that is
asymptotically AdS. These are obtained by imposing additional
constraints on the $F(R)$ model parameters. We have investigated
their asymptotic behavior, computed their masses, and obtained
constraints on the parameters to ensure regularity of the metric.
Our investigation of Dolgov-Kawasaki stability of these solutions
indicates that their objects are stable.

It is possible to obtain more solution sets for Eguchi-Hanson like
(\ref{E-Hlike}) space-times using our approach. However an inspection
of their basic properties indicates that they  are not physical
(see appendix for more details).

It would be worthwhile to investigate the thermodynamic as well as
dynamical stability of  solutions we have found. Other
interesting questions, such as whether or not the sub-class of
theories yielding the EH-like metrics we have found obey
Birkhoff's theorem \cite{Oliva}, remain interesting problems for
future consideration.

\begin{acknowledgements}
This work has been supported financially by Research Institute for
Astronomy \& Astrophysics of Maragha (RIAAM) under research
project No. 1/2348.
The work of R. B. Mann has been supported by the Natural
Science and Engineering Research Council of Canada.
\end{acknowledgements}
\newpage

\begin{center}
\textbf{APPENDIX}
\end{center}
\textbf{A2: Fourth and fifth solution sets:}

We produce here two solution sets with the same limitation on the free
parameters. Considering the model parameters in the third solution set, one can
find two additional solution sets of the field equation (\ref{FE}) with
metric function (\ref{E-Hlike}) as follows
\begin{eqnarray}
IV &:&\left\{
\begin{array}{l}
g(r)=1-Cr, \\
h(r)=\frac{2\chi }{41C}r-\frac{2\left( 41C^{2}-12\chi \right) }{533C^{2}}-%
\frac{6\left( 205C^{2}-8\chi \right) }{2665C^{3}}\left( \frac{2}{r}+\frac{1}{%
Cr^{2}}\right) ,
\end{array}
\right.  \label{FourthSol} \\
V &:&\left\{
\begin{array}{l}
g(r)=\frac{2\chi }{41C}r+\frac{5\chi }{82C^{2}}-\frac{8\left( 82C^{2}-5\chi
\right) }{451C^{3}}\left( \frac{1}{r}+\frac{3}{Cr^{2}}\right) , \\
h(r)=1-Cr,
\end{array}
\right.  \label{FifthSol}
\end{eqnarray}
where $\chi $ and $C$ are integration constant. Since $n=0$, and as we noted
for third solution set, one can redefine $\kappa-2\Lambda $ as an effective
cosmological constant.

Calculating the Kretschmann scalar leads to the following asymptotics
\begin{eqnarray}
\lim_{r\longrightarrow 0}R_{\alpha \beta \mu \nu }R^{\alpha \beta \mu \nu }
&\longrightarrow &\infty ,  \nonumber \\
\lim_{r\longrightarrow \infty }R_{\alpha \beta \mu \nu }R^{\alpha \beta \mu
\nu } &=&\frac{219\chi ^{2}}{1681}.  \label{RR45}
\end{eqnarray}
Considering solution III with negative $C$, one can find that
$h(r)$ flips sign at large $r$ and $g(r)$ doesn't, and so we have
a cosmological horizon. Respectively adjusting $\chi $ to
$\frac{2050C^{2}}{197}$ and $\frac{5248C^{2}}{419}$ in solutions
III and IV, we find that these space-times have a causal horizon
since the functions $g(r)$\ and $h(r)$ both have the same root in
$r=1/C$.

One can note that these kinds of solutions both have the unphysical
asymptotic structure. Here, we compute second derivative of the $F(R)$
function for the mentioned three sets
\begin{equation}
F_{RR}=\frac{-1}{\kappa +\chi -2\Lambda },  \label{FRRthird}
\end{equation}
which confirm that this solution is stable for $\Lambda >\frac{1}{2}\left(
\kappa +\chi \right) $.
\bigskip

\textbf{A3: Sixth and seventh solution sets:}

Looking at Eqs. (\ref{FE}) and (\ref{E-Hlike}), and considering
Eq. (\ref{faIparameters}), one can obviously obtain two solution
sets with constant curvature scalar ($R=\chi $) as follows
\begin{eqnarray}
VI &:&\left\{
\begin{array}{l}
g(r)=1-\frac{C}{6}r^{2}, \\
h(r)=\frac{3\chi -C}{11C}+\frac{54\left( \chi -4C\right)
}{55C^{2}r^{2}},
\end{array}
\right.  \label{SixthSol} \\
VII &:&\left\{
\begin{array}{l}
g(r)=\frac{\chi }{22Cr^{2}}-\frac{4}{5Cr^{4}}, \\
h(r)=1-Cr^{4},
\end{array}
\right.  \label{SeventhSol}
\end{eqnarray}
where $\kappa$ and $n$ are free.

For the sixth solution set (Eq. (\ref{SixthSol})), calculations
show that the Kretschmann scalar diverges at $r=0$. It is more
interesting to note that notwithstanding previous solutions, it
depends on two parameters as $r\longrightarrow \infty $
\begin{equation}
\lim_{r\longrightarrow \infty }R_{\alpha \beta \mu \nu }R^{\alpha
\beta \mu \nu }=\frac{1}{121}\left( 27\chi ^{2}+\frac{56\chi
C}{3}+\frac{416C^{2}}{3} \right).
\end{equation}
Hence there is a curvature singularity at $r=0$. This singularity will be
hidden behind an event horizon provided either
\[
4C>\chi >\frac{C}{3}
\]
provided $C$ and $\chi $ are both positive. However the space-time ends at
the coordinate value $r=\sqrt{6/C}$, and so does not describe a black hole
with familiar asymptotic properties.

For $C$ and $\chi $ both negative there will be an event horizon cloaking
the singularity provided
\[
|\chi |>4|C|
\]
In this case the space-time geometrically consists of a squashed 3-sphere
whose squashing parameter grows exponentially with $r$.

The seventh solution set (Eq. (\ref{SeventhSol})) has a
curvature singularity at $r=0$; for large $r$, we obtain
\begin{equation}
\lim_{r\longrightarrow \infty }R_{\alpha \beta \mu \nu }R^{\alpha \beta \mu
\nu }=\frac{27\chi ^{2}}{121}.
\end{equation}
Note that the circle described by $\psi $ is growing
exponentially relative to the 2-sphere as $r\rightarrow \infty $ and so
these solutions both have the same types of problems as the instanton
solutions. In order to check the Dolgov-Kawasaki stability, we should obtain
$F_{RR}$. It is easy to show that $F_{RR}$ is the same as Eq. (\ref{EHFRR}).

\end{document}